\documentclass[10pt,english]{article}
\usepackage{times}
\usepackage[T1]{fontenc}
\usepackage[latin1]{inputenc}
\usepackage{amsmath}
\usepackage{amssymb}

\makeatletter

\newcommand{\noun}[1]{\textsc{#1}}

\usepackage{slashed}

\usepackage{babel}
\makeatother
\begin{document}

\title{\textbf{Conserving the lepton number $L_{e}-L_{\mu}-L_{\tau}$ in
the exact solution of a 3-3-1 gauge model with right-handed neutrinos }}

\author{\noun{ADRIAN} PALCU}

\date{\emph{Department of Theoretical and Computational Physics - West
University of Timi\c{s}oara, V. P\^{a}rvan Ave. 4, RO - 300223 Romania}}

\maketitle
\begin{abstract}
In this paper we consider a plausible scenario with conserved lepton
number \textbf{$L=L_{e}-L_{\mu}-L_{\tau}$} within the framework of
the exact solution of a particular 3-3-1 gauge model. We discuss the
consequences of conserving this global leptonic symmetry from the
viewpoint of the neutrino mass matrix constructed via special Yukawa
terms (involving tensor products among Higgs triplets). We prove that
the actual experimental data can naturally be reproduced by our scenario
since soft breaking terms with respect to this lepton symmetry are
properly introduced. As a consequence, our solution predicts for the
neutrino sector the correct mass splitting ratio ($\Delta m_{12}^{2}/\Delta m_{23}^{2}\simeq0.033$),
the inverted mass hierarchy, the correct values for the observed mixing
angles ($\sin^{2}\theta_{23}\simeq0.5$ and $\sin^{2}\theta_{12}=0.31$)
and the absolute mass of the lightest neutrino ($m_{0}\sim0.001$eV)
independent of the breaking scale of the model.


\end{abstract}

\section{Introduction}

In a recent series of papers (Refs. \cite{key-1,key-2,key-3}), the
author has developed an original method for contructing the neutrino
mass matrix within the framework of a particular 3-3-1 gauge model
with no other restrictive additional symmetries (such as the lepton
number). A proper tensor product among the Higgs triplets - designed
to recover (successively the spontaneous symmetry breakdown) the well
known mass-generating Yukawa terms in the unitary gauge - is exploited
within the exact solution of a model based on the gauge group $SU(3)_{c}\otimes SU(3)_{L}\otimes U(1)_{X}$
that does not contain particles with exotic electric charges. Here
we mean by the exact solution (see for further details the general
prescriptions in Ref. \cite{key-4}) a specific algebraical method
relying on an appropriate parametrization in the scalar sector of
the model. It provides us with the exact mass eigenstates and mass
eigenvalues for the gauge bosons and the charges (both the electric
and neutral ones) of all the involved particles, just by exactly solving
certain equations. 

The charged fermions acquire their masses through traditional Yukawa
couplings. At the same time, they essentially determine the texture
of the neutrino mass matrix, since the same coupling coefficient acts
for both the charged lepton and its neutrino partner. All these results
can be achieved \cite{key-1} just by tuning a sole free remaining
parameter in the model. Regarding the neutrino sector, the predictions
claimed by this method include \cite{key-2}: the inverted mass hierarchy,
the bi-maximal mixing and - for the first time in the literature,
as we know - the minimal absolute value in the neutrino mass spectrum
$m_{0}\simeq0.0035$eV. 

The price of dealing with only one free parameter resided in a very
large breaking scale of the model and, consequently, in some very
heavy new gauge bosons that had largely overtaken their lower experimental
mass limit \cite{key-5} (arround $1$TeV for non-SM bosons). In order
to improve the phenomenological consequences regarding the breaking
scale, a new approach was proposed: a canonical seesaw mechanism can
arise in the model, just by altering the parameter matrix of the scalar
sector with a small amount (a fine-tuning parameter) \cite{key-3}.
This second parameter allows for decoupling the neutrino phenomenology
from the breaking scale issue (generating thus reasonable masses in
the boson sector), while keeping unaltered all the neutrino masses
and mixing angles achieved in the one-parameter version \cite{key-2}.

Here we intend to improve the original one-parameter solution of the
3-3-1 model of interest (briefly presentetd in Sec. 2) in a different
way. This time, a new symmetry is taken into consideration, namely
the global lepton number $L=L_{e}-L_{\mu}-L_{\tau}$ (Sec. 3) and
its implications for our model are discussed. Then, some new small
parameters ($\alpha,\beta,\gamma$) are introduced in the special
Yukawa terms in order to softly break this global lepton symmetry.
When the $\mu-\tau$ interchange symmetry is invoked the parameters
get a particular ratio, namely $\gamma/\beta=m(\mu)/(\tau)$. Under
these circumstances, the diagonalization of the mass matrix leads
to predictions in good agreement with the experimental values for
mass splitting ratio and mixing angles. We also consider as a big
success of our scenario the minimal absolute neutrino mass computed
independent of any parameter, depending only on the precise account
for the mixing angles. A few remarks concerning phenomenological aspects
of our method are sketched in the final section (Sec. 4).

\section{Preliminaries}

In order to get to the very essence of the resulting phenomenology
of introducing the global lepton symmetry $L=L_{e}-L_{\mu}-L_{\tau}$
into the 3-3-1 model without exotic electric charges, we start by
presenting the particle content of this model and the original manner
to generate neutrino mass matrix via special Yukawa terms.

\subsection{Fermion content of the model}

The femion sector of the pure left (subscript $L$) 3-3-1 gauge model
with right-handed neutrinos (namely model D in Ref. \cite{key-6})
consists of two distinct sectors: lepton families and quark families.
All the lepton generations obey the same representation with respect
to the gauge group:

\begin{equation}
\begin{array}{ccccc}
f_{\alpha L}=\left(\begin{array}{c}
e_{\alpha}\\
\nu_{\alpha}\\
\nu_{\alpha}^{c}\end{array}\right)_{L}\sim(\mathbf{1,3^{\mathbf{\mathbf{*}}}},-1/3) &  &  &  & e_{\alpha L}^{c}\sim(\mathbf{1},\mathbf{1},1)\end{array}\label{Eq.1}\end{equation}
where $e_{\alpha}=e,\mu,\tau$ are the well known charged leptons
from SM, while $\nu_{\alpha}=\nu_{e},\nu_{\mu},\nu_{\tau}$ are their
neutrino partners. We mean by superscript $c$ the charge conjugation
carried out by the operator $C=i\gamma^{2}\gamma^{0}$ (for more details
we reffer the reader to Appendix B in Ref. \cite{key-4}).

The \emph{}quarks come (according to the anomaly cancellation requirement
that preserves the renormalizability) in three distinct generations
as follows: 

\begin{equation}
\begin{array}{ccc}
Q_{iL}=\left(\begin{array}{c}
u_{i}\\
d_{i}\\
D_{i}\end{array}\right)_{L}\sim(\mathbf{3,3},0) &  & Q=\left(\begin{array}{c}
d\\
u\\
U\end{array}\right)_{L}\sim(\mathbf{3},\mathbf{3}^{\mathbf{*}},1/3)\end{array}\label{Eq.2}\end{equation}

\begin{equation}
\begin{array}{ccc}
(d_{L})^{c},(d_{iL})^{c}\sim(\mathbf{3^{\mathbf{*}}},\mathbf{1},1/3) &  & (u_{L})^{c},(u_{iL})^{c}\sim(\mathbf{3^{\mathbf{*}}},\mathbf{1},-2/3)\end{array}\label{Eq.3}\end{equation}

\begin{equation}
\begin{array}{ccccccccc}
(U_{L})^{c}\sim(\mathbf{3^{\mathbf{*}},1},-2/3) &  &  &  &  &  &  &  & (D_{iL})^{c}\sim(\mathbf{3^{\mathbf{*}},1},1/3)\end{array}\label{Eq.4}\end{equation}
with $i=1,2$ and capital letters denoting exotic quarks. The numbers
in brackets following the left-handed fermion felds (Eqs. (1) - (4))
label the representations and their characters with respect to the
gauge group $SU(3)_{c}\otimes SU(3)_{L}\otimes U(1)_{X}$. Note that
the third quark family has to obey a different representation compared
to the other two families. 

The main phenomenology of the exact solution of this model can be
found in Ref. \cite{key-1} where a special generalized Weinberg transformation
is also given in order to separate the neutral bosons ($A_{\mu}$-
the electromagnetic one, $Z_{\mu}$- the Weinberg boson from Standard
Model, and $Z_{\mu}^{\prime}$- the new neutral boson of the present
theory) that get their exact eigenstates. Here we are concerned only
with the neutrino mass issue since some special coupling terms are
introduced in the Yukawa sector and a global symmetry is added as
well.

\subsection{The resulting neutrino mass matrix}

The masses of the fermions in the model are generated by the Yukawa
Lagrangian. In our approach \cite{key-1} it looks like:

\begin{equation}
G_{\alpha\beta}\bar{f}_{\alpha L}(\phi^{(\rho)}e_{\beta L}^{c}+Sf_{\beta L}^{c})+h.c.\label{Eq.5}\end{equation}
with $S=\phi^{-1}\left(\phi^{(\eta)}\otimes\phi^{(\chi)}+\phi^{(\chi)}\otimes\phi^{(\eta)}\right)\sim(\mathbf{1},\mathbf{6},-2/3)$
and $G_{\alpha\beta}$ as the coupling coefficients of the lepton
sector. The Yukawa sector relies on the scalar triplets $\left\{ \phi^{(\rho)}\sim(\mathbf{1},\mathbf{3^{*}},2/3),\phi^{(\chi)},\phi^{(\eta)}\sim(\mathbf{1},\mathbf{3^{*}},-1/3)\right\} $
(see for details of constructing the Higgs sector in 3-3-1 models
with no exotic electric charges Ref. \cite{key-1}). We recall that
the exact solution for the 3-3-1 model of interest here leads to the
one-parameter matrix $\eta=\left(1-\eta_{0}^{2}\right)diag\left[a/2\cos^{2}\theta_{W},1-a,a\left(1-\tan^{2}\theta_{W}\right)/2\right]$
which determines the VEVs alignment in the Higgs sector $\left\langle \phi^{(i)}\right\rangle =\eta^{i}\left\langle \phi\right\rangle ,i=1,2,3$
(successively the SSB), according to the general prescriptions of
the method shown in Sec. 4.1 of Ref. \cite{key-4}. Obviously, $\theta_{W}$
is the Weinberg angle from SM. 

Note that the Higgs field $\phi$ plays the role of the ''norm''
for the geometrized scalar sector and it is therefore a main factor
in all the VEVs. It is appropriate for one to consider it as the homologue
of the neutral scalar field of the Standard Model, since their vacuum
expectation values supply the masses for the particle of the model
they act in.

The charged leptons get their masses via traditional Yukawa couplings
as one can easily observe in Eq.(5): $m(e)=A\left\langle \phi^{(\rho)}\right\rangle $,
$m(\mu)=B\left\langle \phi^{(\rho)}\right\rangle $, $m(\tau)=C\left\langle \phi^{(\rho)}\right\rangle $.
Obviously, $G_{ee}=A$, $G_{\mu\mu}=B$, $G_{\tau\tau}=C$, but we
specify in advance that our notations include: $G_{e\mu}=D$, $G_{e\tau}=E$,
$G_{\mu\tau}=F$ for the off-diagonal terms in the flavor basis. 

Neutrino mixing is expressed by $\nu_{\alpha L}(x)=\sum_{i=1}^{3}U_{\alpha i}\nu_{iL}(x)$,
where $\alpha=e,\mu,\nu$ label the flavor space (flavor eigenstates)
while $i=1,2,3$ denote the massive physical eigenstates. We consider
throughout the paper the physical neutrinos as Majorana fields, \emph{i.e.}
$\nu_{iL}^{c}(x)=\nu_{iL}(x)$. The neutrino mass term in the Yukawa
sector yields then: 

\begin{equation}
\mathcal{-L}_{Y}=\frac{1}{2}\bar{\nu}_{L}M\nu_{L}^{c}+H.c\label{Eq. 6}\end{equation}
with $\nu_{L}=\left(\begin{array}{ccc}
\nu_{e} & \nu_{\mu} & \nu_{\tau}\end{array}\right)_{L}^{T}$ where the superscript $T$ stands for ''transposed'' . The mixing
matrix $U$ that diagonalizes the mass matrix in the manner $U^{+}MU=m_{ij}\delta_{j}$
has in the standard parametrization the form: 

\begin{equation}
U=\left(\begin{array}{ccc}
c_{2}c_{3} & s_{2}c_{3} & s_{3}e^{-i\delta}\\
-s_{2}c_{1}-c_{2}s_{1}s_{3}e^{i\delta} & c_{1}c_{2}-s_{2}s_{3}s_{1}e^{i\delta} & c_{3}s_{1}\\
s_{2}s_{1}-c_{2}c_{1}s_{3}e^{i\delta} & -s_{1}c_{2}-s_{2}s_{3}c_{1}e^{i\delta} & c_{3}c_{1}\end{array}\right)\label{Eq. 7}\end{equation}
with natural substitutions: $\sin\theta_{23}=s_{1}$, $\sin\theta_{12}=s_{2}$,
$\sin\theta_{13}=s_{3}$, $\cos\theta_{23}=c_{1}$, $\cos\theta_{12}=c_{2}$,
$\cos\theta_{13}=c_{3}$ for the mixing angles, and $\delta$ for
the CP phase. 

Our procedure provides (after the SSB) the following new symmetric
mass matrix for the neutrinos involved in the model:

\begin{equation}
M=4\left(\begin{array}{ccc}
A & D & E\\
D & B & F\\
E & F & C\end{array}\right)\frac{\left\langle \phi^{(\eta)}\right\rangle \left\langle \phi^{(\chi)}\right\rangle }{\left\langle \phi\right\rangle }\label{Eq. 8}\end{equation}

The next task is to diagonalize the matrix (8) in order to get the
physical eigenstates of the massive neutrinos. This procedure will
lead (within the Case 1 in Ref. \cite{key-1}, which is the only acceptable
one from the phenomenological point of view) to the following generic
solution:

\begin{equation}
m_{i}=f_{i}\left[\theta_{12},\theta_{23},\theta_{13},m(e),m(\mu),m(\tau)\right]\left(\frac{2a}{\sqrt{1-a}}\right)\frac{\sqrt{1-2\sin^{2}\theta_{W}}}{\cos^{2}\theta_{W}}\label{Eq. 9}\end{equation}
with $i=1,2,3$. In these expressions $f_{i}s$ are analytical functions
depending on the mixing angles and the charged lepton masses in a
particular way to be determined.

\section{Conserving the lepton number $L=L_{e}-L_{\mu}-L_{\tau}$}

In the following we assume that the lepton number $L=L_{e}-L_{\mu}-L_{\tau}$
is conserved (or approximately conserved, as suggested by a lot of
papers concerning this issue \cite{key-7} - \cite{key-21}) in the
3-3-1 model presented above. We first analyze the consequences of
this global symmetry for the neutrino sector and then the modifications
provided by soft-breaking terms with respect to this symmetry. The
modifications occur in the mass matrix due to some new small parameters
introduced in the Yukawa sector.

\subsection{Neutrino mass spectrum}

At this point, if one imposes for the lepton sector an additional
global symmetry given by the lepton number $L=L_{e}-L_{\mu}-L_{\tau},$
certain coupling coefficients in the Yukawa Lagrangian get suppresed,
namely $A=B=C=F=0$, since the following assignement for the lepton
number holds: $L(f_{eL})=L(e_{R})=1$, $L(f_{\mu L})=L(\mu_{R})=-1$
and $L(f_{\tau L})=L(\tau_{R})=-1$, while all the scalar fields carry
zero lepton number. 

Hence, the mass matrix with the exact global $L$ symmetry becomes:

\begin{equation}
M=\left(\begin{array}{ccc}
0 & D & E\\
D & 0 & 0\\
E & 0 & 0\end{array}\right)\left(\frac{2a}{\sqrt{1-a}}\right)\frac{\sqrt{1-2\sin^{2}\theta_{W}}}{\cos^{2}\theta_{W}}\left\langle \phi\right\rangle \label{Eq.10}\end{equation}

The concrete forms of $f_{i}s$ remain to be computed by solving the
following set of equations:

\begin{equation}
\begin{cases}
\begin{array}{c}
f_{1}=-2c_{1}c_{2}s_{2}D+2s_{1}c_{2}s_{2}E\\
0=-c_{1}s_{2}^{2}D+c_{1}s_{2}^{2}D-s_{1}c_{2}^{2}E+s_{1}s_{2}^{2}E\\
0=c_{2}s_{1}{}D+c_{2}c_{1}E{}\\
0=c_{1}c_{2}^{2}D-c_{1}s_{2}^{2}D+s_{1}s_{2}^{2}E-s_{1}c_{2}^{2}E\\
f_{2}=2c_{1}c_{2}s_{2}D-2s_{1}c_{2}s_{2}E\\
0=s_{1}s_{2}{}D+c_{1}s_{2}E{}\\
0=s_{1}c_{2}{}D+c_{1}c_{2}E{}\\
0=s_{1}s_{2}{}D+c_{1}s_{2}E{}\\
f_{3}={}0\end{array}\end{cases}\label{Eq.11}\end{equation}
obtained straightforwardly from Eq. (10) via $U^{+}MU=diag(m_{1},m_{2},m_{3})$.
The lines 3, 6, 7 and 8 in Eqs. (11) express the same condition, namely:
$D=-E\cot\theta_{23}$. The lines 2 and 4 in the set of equations
(11) are fulfiled simultaneously if and only if $\cos^{2}\theta_{12}=\sin^{2}\theta_{12}$
(maximal solar mixing angle). 

Under these circumstances, taking into consideration the maximal atmospheric
mixing angle too, the solution reads:

\begin{equation}
m_{1}=\left|m_{2}\right|=\sqrt{2}D\left(\frac{2a}{\sqrt{1-a}}\right)\frac{\sqrt{1-2\sin^{2}\theta_{W}}}{\cos^{2}\theta_{W}}\left\langle \phi\right\rangle \label{Eq.12}\end{equation}

\begin{equation}
m_{3}=0\label{Eq.13}\end{equation}
giving rise to a $\mu-\tau$ interchange symmetry \cite{key-22} -
\cite{key-30}.

\subsection{Introducing soft breaking terms}

We exploit in the following the consequences of assuming some soft
breaking terms with respect to the global lepton number $L$, by introducing
the free parameters $\alpha$, $\beta$, $\gamma$ in the Yukawa Langrangian:

\begin{equation}
\begin{cases}
\begin{array}{c}
G_{ee}\bar{f}_{eL}(\phi^{(\rho)}e_{L}^{c}+\alpha Sf_{eL}^{c})+h.c.\\
G_{\mu\mu}\bar{f}_{\mu L}(\phi^{(\rho)}\mu_{L}^{c}+\beta Sf_{\mu L}^{c})+h.c.\\
G_{\tau\tau}\bar{f}_{\tau L}(\phi^{(\rho)}\tau_{L}^{c}+\gamma Sf_{\tau L}^{c})+h.c.\end{array}\end{cases}\label{Eq. 14}\end{equation}
Obviously, $\alpha$, $\beta$, $\gamma$ $\in$$[0,1)$. A small
nonzero $F=\beta G_{\mu\tau}$ coupling coefficient is also allowed.
Now, the mass matrix of the neutrino sector stands:

\begin{equation}
M=\left(\begin{array}{ccc}
\alpha A & D & D\\
D & \beta B & \beta F\\
D & \beta F & \gamma C\end{array}\right)\left(\frac{2a}{\sqrt{1-a}}\right)\frac{\sqrt{1-2\sin^{2}\theta_{W}}}{\cos^{2}\theta_{W}}\left\langle \phi\right\rangle \label{Eq.15}\end{equation}

In order to remain in the spirit of the lepton symmetry $L=L_{e}-L_{\mu}-L_{\tau}$
which also favors the interchange symmetry $\mu-\tau$, one can approximate:

\begin{equation}
\beta B\cong\gamma C\label{Eq.16}\end{equation}

Hence, one of the three new parameters has disappeared. The two remaining
ones finally will have to fulfil a particular ratio in order to be
compatible with the phenomenological data, so one remains with only
one parameter to be tuned apart from the main one ($a$ in our model). 

The new resulting mass matrix

\begin{equation}
M(\nu)=\left(\begin{array}{ccc}
\alpha A & D & D\\
D & \beta B & \beta F\\
D & \beta F & \beta B\end{array}\right)\left(\frac{2a}{\sqrt{1-a}}\right)\frac{\sqrt{1-2\sin^{2}\theta_{W}}}{\cos^{2}\theta_{W}}\left\langle \phi\right\rangle \label{Eq.17}\end{equation}
is now to be diagonalized. This task was already carried out in Ref.
\cite{key-2} for the general case, now we just have to insert the
new diagonal entries in the solution obtained there (Eqs. (13) in
Ref. \cite{key-2}) and to impose the maximal atmospheric mixing angle
at least in the numerators, while de denominators will be computed
in the final stage of solving the mass issue by inserting a suitable
approximation for the maximal atmospheric mixing angle. Hence, an
interesting mass split is obtained: \[
f_{1}=-\beta B\frac{\sin^{2}\theta_{12}}{\left(1-2\sin^{2}\theta_{23}\right)\left(1-2\sin^{2}\theta_{12}\right)}+\alpha A\frac{\left(1-\sin^{2}\theta_{12}\right)}{\left(1-2\sin^{2}\theta_{12}\right)},\]

\[
f_{2}=\beta B\frac{\sin^{2}\theta_{12}}{\left(1-2\sin^{2}\theta_{23}\right)\left(1-2\sin^{2}\theta_{12}\right)}+\alpha A\frac{\sin^{2}\theta_{12}}{\left(1-2\sin^{2}\theta_{12}\right)},\]

\begin{equation}
f_{3}=\beta B.\label{Eq.18}\end{equation}

Evidently, it is required a $\gamma^{5}$ transformation performed
on the first neutrino field in order to get the sign change for its
mass ($m_{1}$). 

The mass spectrum in the neutrino sector becomes:

\[
\left|m_{1}\right|=\left[\frac{\beta m(\mu)\sin^{2}\theta_{12}}{\left(1-2\sin^{2}\theta_{23}\right)\left(1-2\sin^{2}\theta_{12}\right)}-\frac{\alpha m(e)\left(1-\sin^{2}\theta_{12}\right)}{\left(1-2\sin^{2}\theta_{12}\right)}\right]\left(\frac{2a}{\sqrt{1-a}}\right)\frac{\sqrt{1-2\sin^{2}\theta_{W}}}{\cos^{2}\theta_{W}},\]

\[
m_{2}=\left[\frac{\beta m(\mu)\sin^{2}\theta_{12}}{\left(1-2\sin^{2}\theta_{23}\right)\left(1-2\sin^{2}\theta_{12}\right)}+\frac{\alpha m(e)\sin^{2}\theta_{12}}{\left(1-2\sin^{2}\theta_{12}\right)}\right]\left(\frac{2a}{\sqrt{1-a}}\right)\frac{\sqrt{1-2\sin^{2}\theta_{W}}}{\cos^{2}\theta_{W}},\]

\begin{equation}
m_{3}=\beta m(\mu)\left(\frac{2a}{\sqrt{1-a}}\right)\frac{\sqrt{1-2\sin^{2}\theta_{W}}}{\cos^{2}\theta_{W}}\label{Eq. 19}\end{equation}

\subsection{Phenomenological predictions}

The physical relevant magnitudes for the neutrino oscillations are
the mass squared differences for solar and atmospheric neutrinos,
defined as: $\Delta m_{12}^{2}=m_{2}^{2}-m_{1}^{2}$ and $\Delta m_{23}^{2}=m_{3}^{2}-m_{2}^{2}$
. They result from the above expressions (Eqs. (19)):

\begin{equation}
\Delta m_{12}^{2}\cong2\alpha\beta\frac{m(e)m(\mu)\sin^{2}\theta_{12}}{\left(1-2\sin^{2}\theta_{12}\right)^{2}\left(1-2\sin^{2}\theta_{23}\right)}\left(\frac{4a^{2}}{1-a}\right)\left(\frac{1-2\sin^{2}\theta_{W}}{\cos^{4}\theta_{W}}\right)\label{Eq.20}\end{equation}

\begin{equation}
\Delta m_{23}^{2}\cong\beta^{2}\frac{m^{2}(\mu)\sin^{4}\theta_{12}}{\left(1-2\sin^{2}\theta_{12}\right)^{2}\left(1-2\sin^{2}\theta_{23}\right)^{2}}\left(\frac{4a^{2}}{1-a}\right)\left(\frac{1-2\sin^{2}\theta_{W}}{\cos^{4}\theta_{W}}\right)\label{Eq.21}\end{equation}

The mass splitting ratio defined as $r_{\Delta}=\Delta m_{12}^{2}/\Delta m_{23}^{2}$
yields in our scenario:

\begin{equation}
r_{\Delta}=2\left(\frac{\alpha}{\beta}\right)\left(\frac{m(e)}{m(\mu)}\right)\left(\frac{1-2\sin^{2}\theta_{23}}{\sin^{2}\theta_{12}}\right)\label{Eq.22}\end{equation}

As in all the cases where our method of generating neutrino masses
is employed, the mass splitting ratio comes out independently of the
breaking scale in the theory. The latter is determined by the parameter
$a$ which is not involved in formula (22). Once again, we have got
an important conclusion: the breaking scale of the model (and consequently,
the boson mass spectrum) and the neutrino phenomenology do not influence
one another (the same happens in Ref. \cite{key-3}, but a different
strategy based on a seesaw mechanism was employed therein). Therefore,
as soon as new data regarding the precision measurements for the mass
of the new neutral boson (other than the Weinberg boson from SM) are
available, one can establish the parameter $a$. As far as we know
\cite{key-5}, the lower limit is $m(Z^{\prime})\geq1.5$TeV, that
claims in our solution for the 3-3-1 mode:l $a\leq0.06$ and $\left\langle \phi\right\rangle \geq1$TeV
\cite{key-1,key-3}.

The next step is to implement the specific values for the atmospheric
and solar mixing angles into formula (22) and estimate how they can
influence the mass splitting ratio. For instance, if the plausible
$\sin^{2}\theta_{23}=0.499$ and $\sin^{2}\theta_{12}=0.31$ are taken
into account, and assuming the charged lepton masses \cite{key-5}
$m(e)=0.511$ MeV and $m(\mu)=106$MeV, one obtaines $r_{\Delta}\simeq0.033$
(very close to the value supplied by data) if 

\begin{equation}
\frac{\alpha}{\beta}=530.5\label{Eq.23}\end{equation}
This estimate suggests that the added matrix responsable for the soft
violation of the initial conserved $L$ symmetry can be actually considered
as a small perturbation of the form: $\delta M=\varepsilon diag(1,1,1)$.

The method presented above allows one to estimate the sum of the absolute
masses in the neutrino sector: 

\begin{equation}
\sum_{i=1}^{3}m_{i}\simeq\frac{2\beta m(\mu)\sin^{2}\theta_{12}}{\left(1-2\sin^{2}\theta_{12}\right)\left(1-2\sin^{2}\theta_{23}\right)}\left(\frac{2a}{\sqrt{1-a}}\right)\frac{\sqrt{1-2\sin^{2}\theta_{W}}}{\cos^{2}\theta_{W}}\label{Eq.24}\end{equation}

This is experimentally restricted to: $\sum_{i=1}^{3}m_{i}\sim1$eV,
if we take into consideration the Troitsk \cite{key-31} and Mainz
\cite{key-32,key-33} experiments. On the other hand, combining Eqs.
(19) and (23) one obtaines:

\begin{equation}
\sum_{i=1}^{3}m_{i}=\frac{2\sin^{2}\theta_{12}}{\left(1-2\sin^{2}\theta_{12}\right)\left(1-2\sin^{2}\theta_{23}\right)}m_{0}\label{Eq.25}\end{equation}
with minimal neutrino mass $m_{0}=m_{3}$ . This leads (with the above
considered values for mixing angles) to $m_{0}\simeq0.001$eV.

\section{Concluding remarks}

In this paper we have developed a strategy that combines the exact
solution of a particular 3-3-1 gauge model with possible global leptonic
symmetry $L=L_{e}-L_{\mu}-L_{\tau}$. When this additional symmetry
is rigourously conserved, the neutrino mass spectrum exhibits an inverted
hierarchy with two degenerate masses and the third one equal to zero.
At the same time this symmetry restricts the mixing angles to the
bi-maximal setting only. This state of affairs can be naturally overtaken
if one deals with the approximate global leptonic symmetry, by introducing
some small terms that softly violates this symmetry in the Yukawa
sector. The results are amazing: the correct predictions regarding
the mass splitting ratio $\Delta m_{12}^{2}/\Delta m_{23}^{2}\simeq0.033$
and the observed values for the mixing angles $\sin^{2}\theta_{23}\simeq0.5$
and $\sin^{2}\theta_{12}=0.31$ arise independently of the main parameter
$a$ in the 3-3-1 model of interest. At the same time, the minimal
absolute mass in the neutrino sector can be computed based on an exact
formula depending only on the accurate account for the mixing angles.
In our approximation, the scenario leads to the minimal mass in the
neutrino spectrum: $m(\nu_{3})=0.001$eV, but this order of magnitude
can decrease if the atmospheric angle gets closer-to-maximal values. 

All the results regarding the mass squared differences and the mixing
angles are achieved just by tuning a second small parameter, let it
be $\alpha$ or $\beta$. There are many advantages of the present
method over the previous ones followed by the author in Refs. \cite{key-2}
and \cite{key-3} respectively. In the one-parameter case \cite{key-2}
with no lepton number conserved and no $\mu-\tau$ interchange symmetry,
the neutrino phenomenology required a very large breaking scale and
a radiative mechanism was set to supply possible deviations from the
bi-maximal mixing. In Ref. \cite{key-3} a second parameter was introduced
(that time in the Higgs sector) accompanied by seek for a canonical
seesaw interpretation in order to get a considerable autonomy for
the neutrino phenomenology from the breaking scale issue. Here, although
a second parameter is introduced too, there is no need for any seesaw
mechanism. Instead, finally a $\mu-\tau$ interchange symmetry remains.
Along with all the old valuable predictions supplied by the exact
solution of our model, the approach based on this second parameter
explaines the neutrino mass phenomenology and, in addition, successfully
passes the difficult challenge of predicting the correct solar and
atmospheric mixing angles.


\begin{thebibliography}{10}
\bibitem{key-1}A. Palcu, \emph{Mod. Phys. Lett.} \emph{A} \textbf{21}, 1203 (2006).
\bibitem{key-2}A. Palcu, \emph{Mod. Phys. Lett.} \emph{A} \textbf{21}, 2027 (2006). 
\bibitem{key-3}A. Palcu, \emph{Mod. Phys. Lett.} \emph{A} \textbf{21}, 2591 (2006). 
\bibitem{key-4}I. I. Cot\u{a}escu, \emph{Int. J. Mod. Phys.} \emph{A} \textbf{12},
1483 (1997). 
\bibitem{key-5}Particle Data Group (S. Eidelman \emph{et. al.)}, \emph{Phys. Lett.}
\emph{B} \textbf{592}, 1 (2004).
\bibitem{key-6}W. A. Ponce, J. B. Florez and L. A. Sanchez, \emph{Int. J. Mod. Phys.}
\emph{A} \textbf{17}, 643 (2002).
\bibitem{key-7}S. T. Petcov, \emph{Phys. Lett.} \emph{B} \textbf{110}, 245 (1982).
\bibitem{key-8}R. Barbieri, L. Hall, D. Smith, A. Strumia and N.Weiner \emph{JHEP}
\textbf{12}, 017 (1998). 
\bibitem{key-9}A. Joshipura and S. Rindani, \emph{Eur. Phys. J.} \emph{C} \textbf{14},
85 (2000).
\bibitem{key-10}R. N. Mohapatra, A. Perez-Lorenzana and C. A. de S. Pires, \emph{Phys.
Lett.} \emph{B} \textbf{474}, 355 (2000). 
\bibitem{key-11}Q. Shafi and Z. Tavartkiladze, \emph{Phys. Lett.} \emph{B} \textbf{482},
145 \textbf{(}2000\textbf{)}.
\bibitem{key-12}L. Lavoura, \emph{Phys. Rev}. \emph{D} \textbf{62}, 093011 (2000).
\bibitem{key-13}W. Grimus and L. Lavoura, \emph{Phys. Rev}. \emph{D} \textbf{62},
093012 (2000).
\bibitem{key-14}T. Kitabayashi and M. Yassue, \emph{Phys. Rev.} \emph{D} \textbf{63},
095002 (2001).
\bibitem{key-15}R. N. Mohapatra, \emph{Phys. Rev.} \emph{D} \textbf{64}, 091301 (2001).
\bibitem{key-16}S. Babu and R. N. Mohapatra, \emph{Phys. Lett.} \emph{B} \textbf{532},
77 (2002). 
\bibitem{key-17}H. S. Goh, R. N. Mohapatra and S. P. Ng, \emph{Phys. Lett.} \emph{B}
\textbf{542}, 116 (2002). 
\bibitem{key-18}D. A. Dicus and H.-J. He and J. N. Ng, \emph{Phys. Lett.} \emph{B}
\textbf{536}, 83 (2002).
\bibitem{key-19}G. K. Leontaris, J. Rizos and A. Psallidas, \emph{Phys. Lett.} \emph{B}
\textbf{597}, 182 (2004).
\bibitem{key-20}W. Grimus and L. Lavoura, \emph{J. Phys. G} \textbf{31}, 683 (2005). 
\bibitem{key-21}G. Altarelli and R. Franceschini, \emph{JHEP} \textbf{03}, 047 (2006).
\bibitem{key-22}T. Fukuyama and H. Nishiura, hep-ph/9702253.
\bibitem{key-23}R. N. Mohapatra and S. Nussinov, \emph{Phys. Rev.} \emph{D} \textbf{60},
013002 (1999).
\bibitem{key-24}E. Ma and M. Raidal, \emph{Phys.Rev. Lett.} \textbf{87}, 011802 (2001). 
\bibitem{key-25}C. S. Lam, \emph{Phys. Lett. B} \textbf{507}, 214 (2001). 
\bibitem{key-26}W. Grimus and L. Lavoura, \emph{Phys. Lett. B} \textbf{572}, 189 (2003).
\bibitem{key-27}T. Kitabayashi and M. Yasue, \emph{Phys. Rev.} \emph{D} \textbf{67},
015006 (2003). 
\bibitem{key-28}W. Grimus and L. Lavoura, \emph{J. Phys. G} \textbf{30}, 73 (2004). 
\bibitem{key-29}Y. Koide, \emph{Phys. Rev.} \emph{D} \textbf{69}, 093001 (2004). 
\bibitem{key-30}A. Ghosal, \emph{Mod. Phys. Lett.} \emph{A} \textbf{19}, 2579 (2004).
\bibitem{key-31}V. M. Lobashev et al., \emph{Prog. Part. Nucl. Phys.} \textbf{48},
123 (2002).
\bibitem{key-32}C. Weinheimer et al., \emph{Nucl. Phys. Proc. Suppl.} \textbf{118},
279 (2003). 
\bibitem{key-33}Ch. Krauss et al., \emph{Eur. Phys. J.} \emph{C} \textbf{40}, 447
(2005). \end{thebibliography}
\end{document}